\documentclass[11pt,a4paper]{article}

\pdfoutput=1
\usepackage{jheppub}

\usepackage{braket}

\definecolor{darkgreen}{RGB}{23,130,10}

\begin{document}
\title{Shrinking the Warm Little Inflaton} 
\date{\today}
\author[a, b]{Paulo B. Ferraz}
\emailAdd{paulo.ferraz@student.uc.pt}

\author[a]{Jo\~{a}o G.~Rosa}
\emailAdd{jgrosa@uc.pt}

\affiliation[a]{Univ. Coimbra, Faculdade de Ciências e Tecnologia da Universidade
de Coimbra and CFisUC, Rua Larga, 3004-516 Coimbra, Portugal}
\affiliation[b]{Departamento de F\'{\i}sica Te\'orica y del Cosmos, Universidad de Granada, Granada-18071, Spain}



\abstract{
We show that warm inflation can be successfully realized in the high temperature regime through dissipative interactions between the inflaton and a single fermionic degree of freedom, provided that the latter's mass is an oscillatory function of the inflaton field value. We demonstrate, in particular, that despite the consequent large amplitude oscillations of the eta slow-roll parameter, their effect is, on average, sufficiently suppressed to allow for a slow-roll trajectory. In addition, we demonstrate that, even though this also induces a parametric resonance that amplifies inflaton perturbations, this has a negligible effect on CMB scales in the relevant parametric range. Hence, the ``Warm Little Inflaton'' scenario can be realized with one less fermionic degree of freedom and no need of imposing an additional discrete interchange symmetry.} 

\maketitle


\section{Introduction}

In the current view of modern cosmology, inflation \cite{Guth:1980zm,Linde:1981mu,Albrecht:1982wi,Albrecht:1982mp,Dolgov:1982th,Abbott:1982hn,Linde:1983gd} is the most compelling mechanism to explain the characteristics of the presently observable universe, namely its flatness, isotropy and homogeneity. In conventional models of inflation, the accelerated expansion of the early universe is driven by the inflaton scalar field in a slow-roll trajectory, where it temporarily mimics the effect of a cosmological constant. One of the most appealing features of the inflationary paradigm is that quantum fluctuations of the inflaton field generate a primordial spectrum of curvature perturbations that may provide the seeds for the observed temperature and polarization anisotropies in the Cosmic Microwave Background (CMB) and the Large-Scale Structure of our universe. 

However, a successful realization of inflation within a complete particle physics framework is challenging. In particular, a slow-roll trajectory requires that the inflaton's mass does exceed the Hubble parameter during inflation, $m_\phi^2 \ll H^2$. Since there are, in general, no symmetries that protect scalar masses from large quantum corrections and which remain unbroken during inflation, single-field slow-roll inflation is quite sensitive to the (yet unknown) details of super-planckian physics, leading to the so-called ``eta-problem'' (see e.g.~\cite{Baumann:2014nda}).

Another important issue to address is the ``graceful exit'' from inflation into a radiation-dominated era, during which we know that the cosmological synthesis of light elements took place. This is conventionally thought to occur through a {\it reheating} period at the end of inflation, with potentially an earlier preheating phase of resonant particle production \cite{Kofman:1997yn,Allahverdi:2010xz}. The main problem with this proposal is that it may be extremely difficult to probe, given that at this stage fluctuations on large CMB scales are already frozen beyond the Hubble radius. This could mean that it may prove nearly impossible to ascertain the inflaton's role within the overall particle physics landscape and how it interacts with other particle species.

Warm inflation \cite{Berera:1995ie,Berera:1996nv,Berera:1996fm} may provide an appealing alternative to the more conventional paradigm (to which we will refer as {\it cold inflation}), where interactions between the inflaton scalar field and other fields play a significant role in the slow-roll dynamics itself. In warm inflation, it is assumed that the inflaton interacts with an ambient nearly-thermal radiation bath, which results in fluctuation-dissipation effects that not only help sustain the field's slow-roll dynamics but also prevent an exponential dilution of the radiation through the associate dissipative particle production. These dissipative effects are, to leading order in the adiabatic regime characteristic of the slow-roll phase, described by a dissipative coefficient $\Upsilon$ and the inflaton follows a Langevin-like equation of the form \cite{Berera:1995wh,Berera:1999ws, Berera:2008ar, Bastero-Gil:2009sdq}:
\begin{align}
\ddot{\phi}+(3H+\Upsilon)\dot{\phi}-\frac{1}{a^2}\nabla^2\phi+V'(\phi) = \xi~,
\label{fulldynamics}
\end{align}
where $\xi$ denotes the nearly Gaussian white noise that satisfies the fluctuation-dissipation relation 
\begin{align}
\langle \xi(x',t')\xi(x,t) \rangle = 2\Upsilon Ta^{-3}\delta(x'-x)\delta(t'-t)~,
\end{align}
and $T$ denotes the temperature of the ambient radiation bath.
From the covariant conservation of the total energy-momentum tensor, the energy density of the radiation bath, $\rho_R=C_RT^4$ with $C_R=\pi^2 g_*/30$ for $g_*$ relativistic degrees of freedom, obeys:
\begin{align}
\dot{\rho}_R+4H\rho_R = \Upsilon \dot{\phi}^2,
\end{align}
and it can be shown from first principles that the source term on the right-hand side results from finite-temperature dissipative particle production \cite{Moss:2008lkw}. This term prevents the otherwise exponential dilution of the radiation bath during inflation, with $4H \rho_R\simeq \Upsilon\dot\phi^2$ in the slow-roll regime. The average inflaton field value then follows the slow-roll trajectory given by:
\begin{align}
3H(1+Q)\dot\phi \simeq -V'(\phi)~,
\end{align}
where $Q=\Upsilon/3H$ is the dissipative ratio, provided that generalized slow-roll conditions are satisfied:
\begin{align}
\epsilon_\phi = \frac{M_P^2}{2}\Big(\frac{V'}{V}\Big)^2 < 1+Q,\quad |\eta_\phi| = M_P^2\Big|\frac{V''}{V}\Big|<1+Q~,
\end{align}
where $M_P$ is the reduced Planck mass. As we can see these are relaxed compared to cold inflation scenarios, where one requires $\epsilon_\phi < 1$ and $|\eta_\phi|<1$, thus allowing for less flat potentials if dissipation is sufficiently strong, i.e.~$Q>1$. This then allows a slow-roll trajectory for an inflaton mass $H<m_\phi<Q^{1/2}H$, thus eliminating or at least alleviating the eta-problem \cite{Berera:1999ws, Berera:2004vm, BasteroGil:2009ec}.

One of the most interesting features of warm inflation is the possibility of a smooth transition between inflation and a radiation-dominated era, with no need for a separate reheating phase \cite{Berera:1996fm}. In particular, the ratio between the radiation and inflaton energy densities during inflation is given by:
\begin{align}
\frac{\rho_R}{V(\phi)}\simeq \frac{1}{2}\frac{Q}{1+Q}\frac{\epsilon_\phi}{1+Q}. \label{radiationoverinflaton}
\end{align}
so that if $Q\gtrsim 1$ at the end of the slow-roll phase when $\epsilon_\phi \simeq 1+Q$, we find $\rho_R\sim \rho_\phi\simeq V(\phi)$ and radiation may then naturally take over as the dominant component.

In addition, the presence of dissipative effects affects the growth of inflaton fluctuations, leading to a unique imprint on the primordial curvature power spectrum that can be used to probe the interactions between the inflaton and the particles present in the ambient cosmic plasma. The inflaton perturbations satisfy the dynamical equation in Fourier space \cite{Berera:1999ws, Graham:2009bf, Hall:2003zp}
\begin{align}
\delta\ddot{\phi}_k+3H(1+Q)\delta\dot{\phi}_k+\Big(\frac{k^2}{a^2}+V_{,\phi\phi}(\phi)\Big)\delta\phi_k=\xi_k,\label{perturbations1}
\end{align}
which leads to a dimensionless curvature power spectrum \cite{Ramos:2013nsa} given by:
\begin{align}
\Delta_{\mathcal{R}}^2=\frac{V_*(1+Q_*)^2}{24\pi^2M_P^4\epsilon_{\phi * }}\left(1+2n_*+\frac{2\sqrt{3}\pi Q_*}{\sqrt{3+4\pi Q_*}}\frac{T_*}{H_*}\right)G(Q_*).
\end{align}
where $n_*$ denotes the inflaton phase space distribution and all quantities are evaluated when the relevant CMB modes become ``superhorizon'' $50-60$ e-folds before inflation ends. The factor $G(Q_*)$ takes into account the dynamical interplay between inflaton and radiation perturbations associated with the temperature-dependence of the dissipation coefficient, and in general needs to be computed numerically \cite{Bastero-Gil:2011rva, Bastero-Gil:2019gao, Montefalcone:2023whx}.

Concrete particle physics implementations of warm inflation are also nevertheless challenging to achieve, with some authors even dubbing it ``impossible'' \cite{Yokoyama:1998ju} in the early days after its original proposal (see also \cite{Berera:1998gx}). Given a Lagrangian density describing the couplings between the inflaton and other fields, one can compute the dissipation coefficient $\Upsilon(\phi, T)$ using standard linear response theory techniques in flat space thermal field theory, provided that $T\gtrsim H$ and that interaction rates within the thermal bath exceed the Hubble rate, $\Gamma\gtrsim H$, to ensure that it remains close to thermal equilibrium despite the disturbances generated by dissipative particle production. 

While these conditions are typically easy to fulfill \cite{Bastero-Gil:2009sdq, Bastero-Gil:2019gao}, particles directly coupled to the inflaton field typically acquire large masses, $m\gtrsim T$, as a result of the large (often super-planckian) field value, as e.g.~for standard inflaton-fermion Yukawa interactions. This means that on-shell dissipative particle production may be Boltzmann-suppressed, and even light particle production mediated by heavy off-shell modes is power-law suppressed at least as $(T/m)^2$ \cite{Berera:2002sp, Bastero-Gil:2010dgy, Bastero-Gil:2012akf}, requiring a very large number of mediator species \cite{Bastero-Gil:2009sdq, Bastero-Gil:2011zxb, Bastero-Gil:2011clw, Bartrum:2012tg, Cerezo:2012ub, Bartrum:2013oka, Bastero-Gil:2013owa}. Even if the high-temperature regime where $T\gtrsim m$ can be realized, one has to worry about the thermal backreaction on the inflaton potential, which may reintroduce the eta-problem through large thermal corrections to the inflaton's mass and which cannot be overcome by the additional dissipative friction.

The ``Warm Little Inflaton'' (WLI) scenario \cite{Bastero-Gil:2016qru} solved these issues for the first time, thus allowing for a simple realization of warm inflation in the high-temperature regime, with moreover a natural embedding in a simple extension of the Standard Model with right-handed neutrinos that also explains light neutrino masses \cite{Levy:2020zfo} (see also \cite{Bastero-Gil:2017wwl, Bastero-Gil:2018uep}). The main idea is to consider interactions between the inflaton and other fields (either fermionic or bosonic \cite{Bastero-Gil:2019gao}) such that the latter's masses are bounded functions of the scalar field value, therefore allowing for $m\lesssim T$ during the slow-roll phase. In addition, a discrete symmetry can be imposed to cancel the leading thermal corrections to the inflaton's mass, thus preventing a reintroduction of the eta-problem that dissipation is supposed to alleviate. This symmetry involves at least two particle species coupled to the inflaton field, and it is exactly this premise (and therefore limitation of the model) that we question in this work.

Anticipating our main results, we will show that warm inflation can be successfully realized by coupling the inflaton field to a {\it single} light fermion field, in such a way that the latter's mass is an oscillatory function of the scalar inflaton value, as follows from a collective symmetry breaking setup. We will show that no eta-problem exists in this case, as opposed to what one may naively expect, and argue that observational predictions are essential identical to the original WLI scenario, within the current level of precision in CMB measurements.

This work is organised as follows. In the next section, we describe our simpler version of the WLI model in terms of the relevant interactions and dynamical quantities, showing that the average inflaton field dynamics is essentially identical to the original WLI model. In section 3, we study the dynamics of inflaton perturbations, in particular exploring the possible development of parametric resonances that could potentially affect the spectrum of primordial curvature perturbations, focusing on how this may or not affect the model's observational predictions. We summarize our main conclusions and discuss prospects for future work motivated by our results in section 4. Throughout this work we consider natural units $k_B=c=\hbar=1$.


\section{Quantum field theory model and inflationary dynamics}

In the original WLI model \cite{Bastero-Gil:2016qru}, the inflaton is a singlet scalar field that results from the collective spontaneous breaking of a U(1) gauge symmetry. The fundamental particle content consists of two complex scalar fields, $\phi_1$ and $\phi_2$, with identical U(1) charge $q$. The model's scalar potential is such that they have the same nonzero vacuum expectation value $\langle \phi_1 \rangle = \langle \phi_2 \rangle = M/\sqrt{2}$ (as enforced by the discrete interchange symmetry as described below). The vacuum manifold may thus be parameterized as:
\begin{align}
\phi_1 = \frac{M}{\sqrt{2}}e^{i(\sigma+\phi)/M}~,\quad\quad\phi_2 = \frac{M}{\sqrt{2}}e^{i(\sigma-\phi)/M}~.
\end{align}
It is easy to check that the overall phase $\sigma(x)$ constitutes the Nambu-Goldstone (NG) boson of the spontaneously broken U(1) gauge symmetry, and in the unitary gauge it can be rotated away in the sense that it is absorbed as the longitudinal component of the now massive U(1) gauge boson. The relative phase $\phi(x)$ remains as a physical scalar degree of freedom in the broken phase, and we note that it is gauge invariant since a local U(1) transformation shifts the phase of the two complex fields by the same amount. Note that the $\sigma(x)$ field inherits a shift symmetry from the underlying U(1) symmetry, which clearly identifies it with the NG degree of freedom.

The U(1) gauge symmetry is therefore consistent with any function $V(\phi)$ that one may add to the original Lagrangian, making this relative phase field an ideal candidate for the inflaton in this setup\footnote{In the original WLI proposal, the inflaton was referred to as a pseudo-NG boson, which may erroneously lead the reader to think of it as a sort of pion or axion, which it is not since there is no associated shift symmetry.}. We may, in fact, take this to be the lightest scalar degree of freedom, since the radial fields' and also the gauge boson's mass can be taken to be $\mathcal{O}(M)\gtrsim H$.

We may also consider interactions between the inflaton and other degrees of freedom consistent with the gauge symmetry. The original model considers two fermion species $\psi_1$ and $\psi_2$ whose left-handed components have U(1) charge $q$, while their right-handed counterparts are neutral. Imposing a discrete interchange symmetry under which $\phi_1\leftrightarrow i\phi_2$ and $\psi_{1L,R} \leftrightarrow \psi_{2L,R}$, one may write the following Yukawa interaction Lagrangian\footnote{This is not the most general Lagrangian as discussed in \cite{Levy:2020zfo} but it is simpler to consider this form without loss of generality.}:
\begin{align}
-\mathcal{L}_{\phi\psi} &= \frac{g}{\sqrt{2}}(\phi_1+\phi_2)\overline{\psi}_{1L}\psi_{1R}-i\frac{g}{\sqrt{2}}(\phi_1-\phi_2)\overline{\psi}_{2L}\psi_{2R}\\
&=gM\cos(\phi/M)\overline{\psi}_{1L}\psi_{1R}+gM\sin(\phi/M)\overline{\psi}_{2L}\psi_{2R}~.
\end{align}
The resulting Dirac masses are therefore bounded oscillatory functions of the inflaton scalar field, $|m_{1,2}|\leq gM$, which can be made light during inflation independently of the inflaton value, provided that $gM\lesssim T \lesssim M$. These interactions lead to thermal corrections to the inflaton potential which, for $m_i\lesssim T,\;i=1,2$ \cite{Kapusta:2006pm, Cline:1996mga}, are of the form:
\begin{align}
V_{T,i} \simeq -\frac{7\pi^2}{180}T^4+\frac{m_i^2T^2}{12}+\frac{m_i^4}{16\pi^2}\Big[\log\Big(\frac{\mu^2}{T^2}\Big)-c_f\Big],
\end{align} 
where $\mu$ is the $\overline{\mathrm{MS}}$ renormalization scale and $c_f \simeq 2.635$. Adding both fermionic contributions, we see that the leading quadratic term is independent of the inflaton, while the parameter $\eta_\phi$ will only receive contributions from the sub-leading Coleman-Weinberg term, which was shown not to lead to an eta-problem on its own \cite{Bastero-Gil:2016qru}.

The dissipation coefficient $\Upsilon$ is the only missing ingredient and it can be computed following the standard techniques in linear response theory \cite{Kapusta:2006pm,Moss:2006gt, Bastero-Gil:2010dgy}. For simplicity, here we only quote the approximate form of the dissipation coefficient, and the reader may look into the details of the computation in \cite{Bastero-Gil:2016qru} and \cite{Levy:2020zfo}:
\begin{align} \label{dissip_coeff}
\Upsilon = C_TT,\quad C_T \simeq \frac{g^2}{h^2}\frac{3}{1-0.34\log h}~,
\end{align}
where it was assumed that interactions in the thermal bath are governed by a Yukawa coupling $h$ coupling the $\psi_{1,2}$ fermions to additional light fermions and scalars. For instance, in the concrete implementation developed in \cite{Levy:2020zfo}, the $\psi_{1,2}$ fermions are right-handed neutrinos, and $h$ corresponds to its coupling to left-handed neutrinos and charged leptons alongside the Higgs field. With this form of the dissipation coefficient, it was shown in \cite{Bastero-Gil:2016qru, Bastero-Gil:2017wwl, Bastero-Gil:2018uep, Levy:2020zfo},  that this model is consistent with Planck CMB data for a quartic inflaton potential (which is excluded within the cold inflation paradigm) and that its embedding within a Standard Model extension naturally leads to the measured light neutrino mass differences. The interchange symmetry protects the inflaton's decay at late times, much like in the proposal of \cite{Bastero-Gil:2015lga}, making it a possible candidate for dark matter \cite{Rosa:2018iff} or quintessential dark energy \cite{Rosa:2019jci}. The model was later adapted to include couplings to scalar rather than fermion fields, in which case the dissipation coefficient is approximately inversely proportional to the temperature \cite{Bastero-Gil:2019gao}. In this case it was possible to realize warm inflation with a quadratic potential in the strong dissipative regime, where the eta-problem is completely solved (while in the fermionic case dissipation can only consistently become strong, i.e. $Q\gtrsim 1$, at the end of inflation).

One may, however, question whether all these ingredients are crucial to obtain a successful realization of warm inflation. The boundedness of the $\psi_{1,2}$ fermion masses is definitely an important issue, given the typically large inflaton field values attained for most (if not all) forms of the inflaton potential $V(\phi)$, often above the Planck scale. Let us, however, eliminate one of the fermion fields, e.g.~$\psi_1$, and consider the form of the resulting finite-temperature corrections to the scalar potential, which to leading order are given by:
\begin{align} \label{thermal_corrections}
V_T \simeq -\frac{7\pi^2}{180}T^4+\frac{g^2M^2}{12}T^2\cos^2(\phi/M)~.
\end{align}
These are obviously sub-leading in the regime where inflation takes place, $\rho_R\sim T^4\ll \rho_\phi$, but the second term leads to an {\it a priori} non-negligible correction to the slow-roll parameters. In particular, we have:
\begin{align}
\Delta\eta_\phi \simeq -\frac{g^2}{18}\left(\frac{T}{H}\right)^2\cos(2\phi/M)~.
\end{align}
Given that $T\gtrsim H$, or otherwise one would be in a cold inflation regime, and that in fact CMB observations favour a regime where $T/H=\mathcal{O}(100)$ and the coupling $g$ is not too suppressed \cite{Bastero-Gil:2016qru, Bastero-Gil:2017wwl, Bastero-Gil:2018uep, Levy:2020zfo}, we see that these corrections may potentially be very large. 

In Fig.~1 we show the results of a numerical solution of the coupled inflaton-radiation system with a dissipation coefficient given by Eq.~(\ref{dissip_coeff}) and a quartic inflaton potential $V(\phi)=\lambda\phi^4$, including the thermal corrections in Eq.~(\ref{thermal_corrections}). We choose parameter values consistent with CMB data for 60 e-folds of inflation after the relevant scales cross the Hubble radius in the original WLI scenario, up to a rescaling of the $g$ coupling by a factor $\sqrt{2}$ to compensate for the $1/2$ reduction of the dissipation coefficient in the absence of the contribution from $\psi_2$. We compare the full solution with that obtained by solving the slow-roll equations
\begin{align} \label{SR_eqs}
\frac{\phi '}{M_P} = -\frac{\sqrt{2\epsilon_\phi}}{1+Q},\quad \frac{Q'}{Q} = \frac{6\epsilon_\phi-2\eta_\phi}{3+5Q},
\end{align}
where primes denote derivatives with respect to the number of e-folds $N_e$, without considering any thermal corrections to the scalar potential.

\begin{figure}[!h]
\centering
\includegraphics[scale = 0.35]{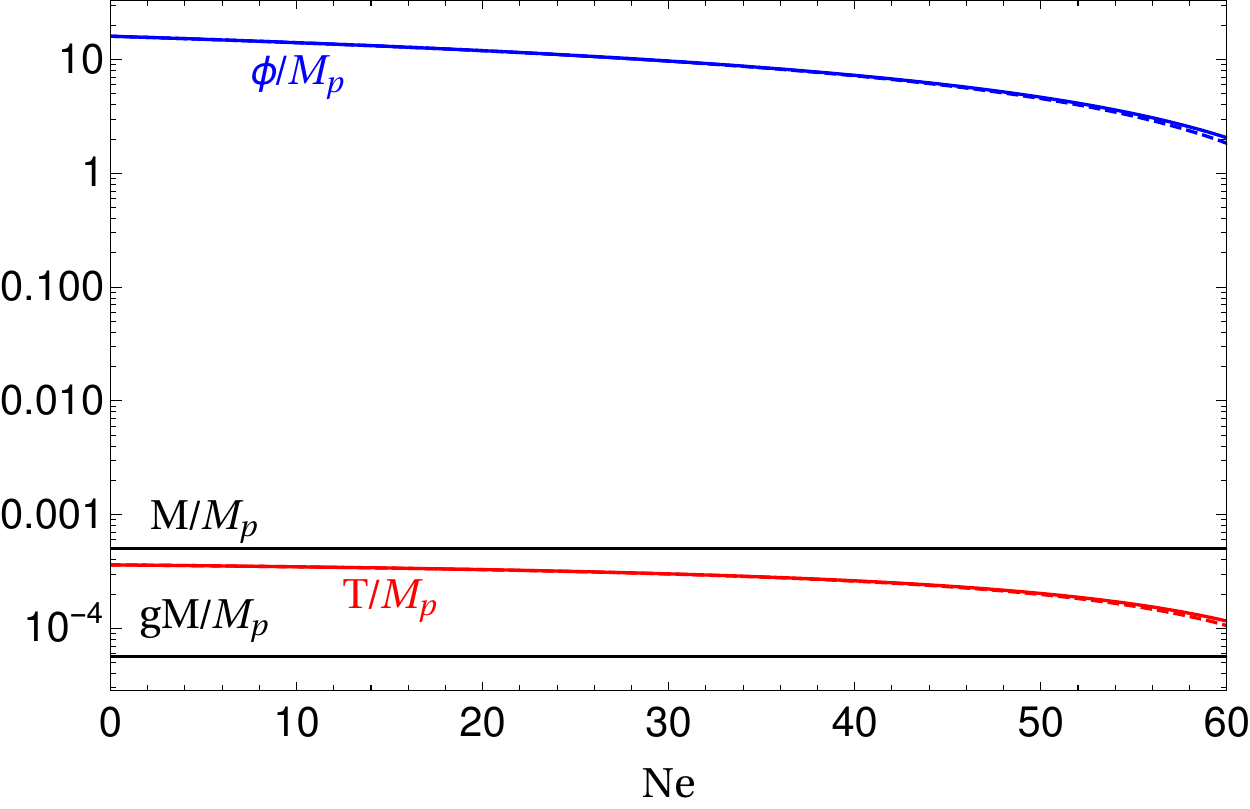}
\includegraphics[scale = 0.35]{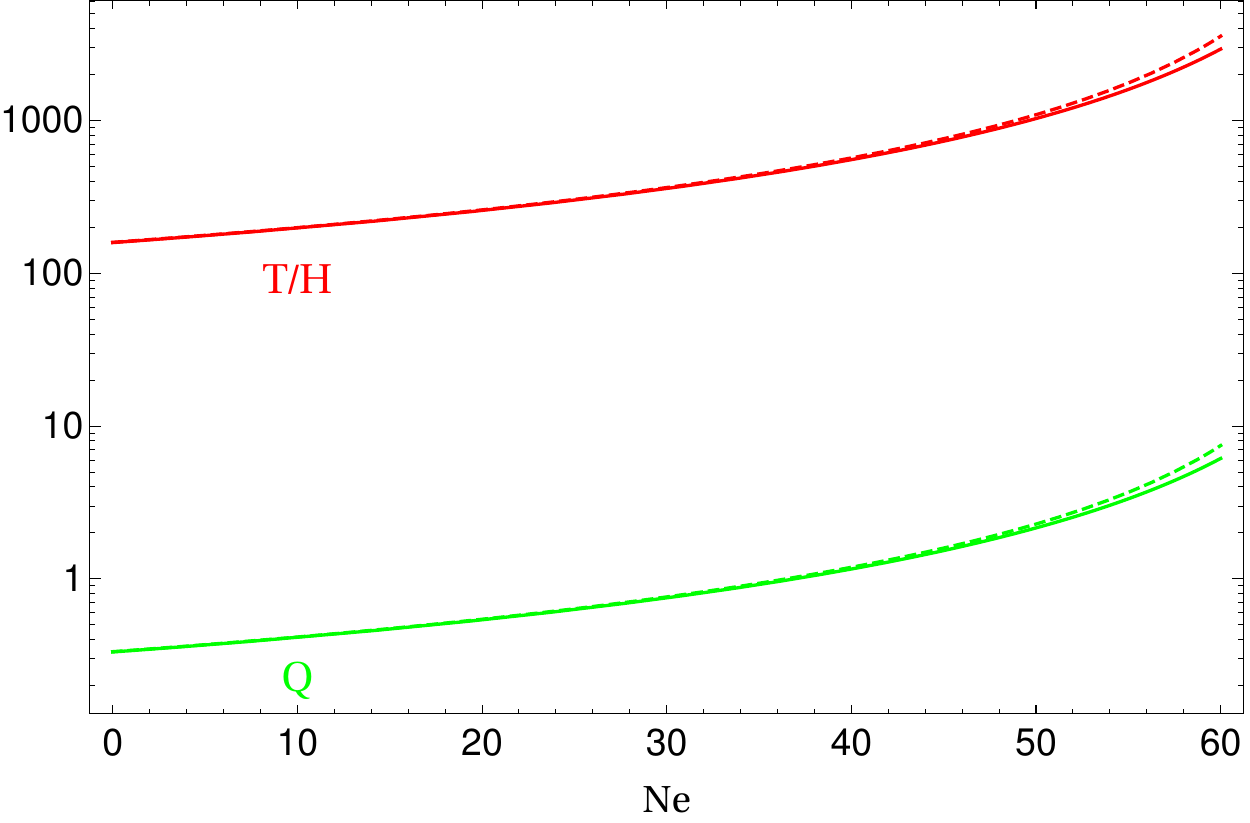}
\caption{Numerical solution (solid lines) for the inflaton-radiation system in the modified WLI scenario with a quartic scalar potential, showing the evolution of the background inflaton field $\phi$, the temperature of the thermal bath $T$, the ratio $T/H$ and the dissipative ratio $Q=\Upsilon/3H$ as a function of the number of e-folds after CMB scales become super-horizon. The evolution of the same quantities is also shown in the slow-roll approximation by the corresponding dashed curves. In this simulation, we considered the values $g\simeq 0.1$ and $h=2$ for the two Yukawa couplings, $g_*=9$ for the number of relativistic degrees of freedom in the radiation bath and $M\simeq 1.2\times10^{15}$ GeV for the U(1) gauge symmetry breaking scale. These match the parameter choices in Fig.~3 of \cite{Bastero-Gil:2016qru}, which yield an observationally consistent scenario in the original WLI setup.}
\label{comparisondyn}
\end{figure}

As one can see in this figure, the full solution follows the slow-roll trajectory for nearly the 60 e-folds of the simulation, with the temperature of the thermal bath below the U(1) gauge symmetry breaking scale $M$ but greater than the fermion's mass, $m_1\sim g M$. The temperature always exceeds the Hubble rate (and one can check that the fermion decay rate as well for the chosen value of the coupling $h$), with inflation ending in the strong dissipation regime that allows for a smooth transition into a radiation-dominated era. Perhaps surprisingly, no oscillatory features are observed throughout the whole dynamical evolution of the field, temperature and dissipative ratio $Q$. However, if we use these results to determine the evolution of the slow-roll parameters, one does find very large amplitude oscillations, as shown in Fig.~2. One would therefore naively expect the slow-roll dynamics to be impossible in this case.

\begin{figure}[!h]
\centering
\includegraphics[scale = 0.35]{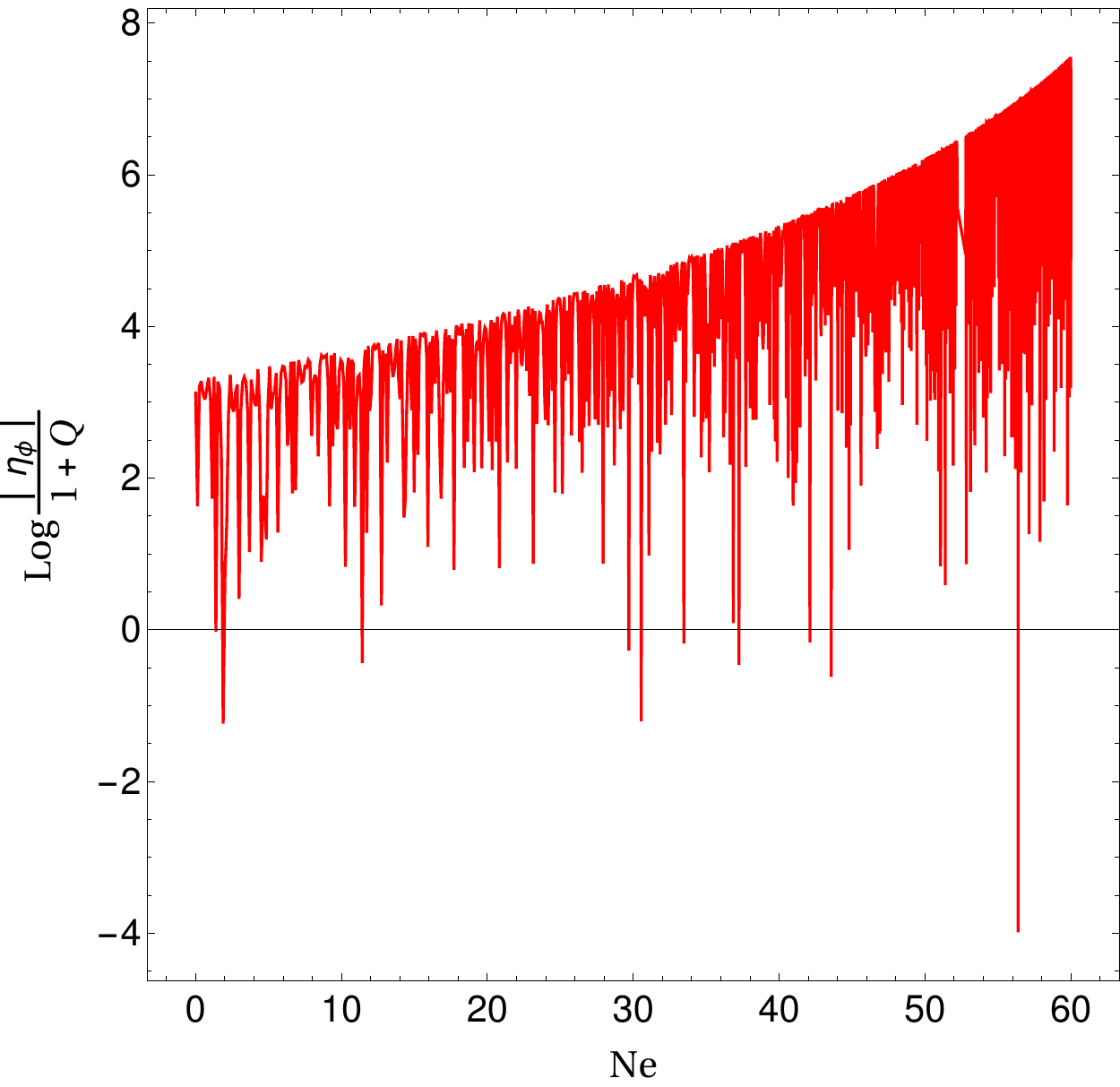}
\includegraphics[scale = 0.35]{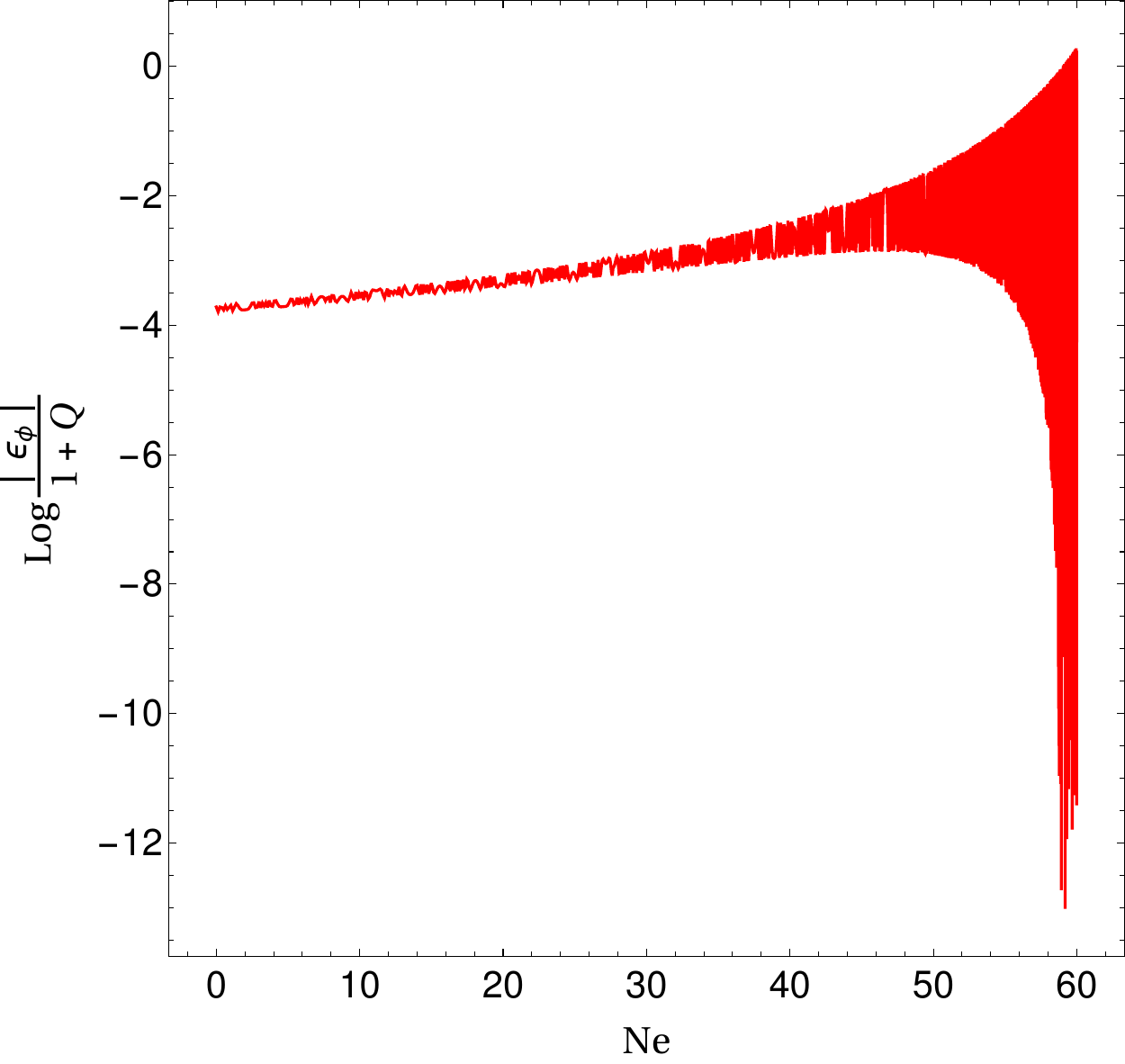}
\caption{Evolution of the slow-roll conditions corresponding to the numerical solution shown in Fig.~1.}
\label{slowrollpara}
\end{figure}
\newpage
The explanation is, in fact, quite simple. As the inflaton field rolls down its potential, albeit slowly, the eta parameter oscillates very quickly between positive and negative values, and only its average effect has an impact on the field's evolution. To better understand this, let us first note that in the slow-roll regime, the field's trajectory is well approximated by $\phi \simeq \phi_i+\dot{\phi}(t-t_i)$ over a period of oscillation of the eta-parameter starting at some arbitrary time $t_i$. We then have:
\begin{align}
\Delta \eta_\phi =-\frac{g^2}{18}\left(\frac{T}{H}\right)^2\cos\left({2\phi_i\over M}+{2\dot{\phi}\over M}(t-t_i)\right) = -\mathcal{A}\cos(\Omega t+\alpha),
\end{align}
where $\mathcal{A} = \frac{g^2}{18}\left(\frac{T}{H}\right)^2$, $\alpha = -2 (\phi_i-\dot{\phi}t_i)/M$ and 
\begin{align}
\Omega = 2{|\dot{\phi}|\over M} \simeq  {2\sqrt{2\epsilon_\phi}\over 1+Q}\left(\frac{M_P}{M}\right)H~.
\end{align}
Note that typically $\Omega\gg H$, so that the eta-parameter oscillates quickly on the Hubble scale. For instance, for the parameters chosen for the numerical solution shown in Fig.~1, we have $\Omega/H\sim 10^3$ already at horizon-crossing of the relevant CMB scales. We may then compute the average correction to the eta-parameter as follows, expanding $\mathcal{A}(t)\simeq \mathcal{A}_i+\dot{\mathcal{A}}(t-t_i)$:
\begin{eqnarray}
\langle \Delta \eta_\phi \rangle &=&\frac{\Omega}{2\pi}\int_{t_i}^{t_i+2\pi/\Omega}\Delta \eta_\phi dt\nonumber\\
&\simeq & -{\Omega\over 2\pi}\dot{\mathcal{A}}\int_{t_i}^{t_i+2\pi/\Omega}(t-t_i)\cos(\Omega t +\alpha)\nonumber\\
&\simeq & -\mathcal{A}{M\over M_P}{Q'\over Q}{1+Q\over \sqrt{2\epsilon_\phi}}\sin\left(2\phi\over M\right)\nonumber\\
&\simeq& -\frac{g^2}{18}\left(\frac{T}{H}\right)^2\left(\frac{M}{M_P}\right)\sqrt{\epsilon_\phi\over2}\sin\left({2\phi\over M}\right)~,\label{ave1}
\end{eqnarray}
where all quantities are evaluated at $t_i$, and in the last line we have used the slow-roll equations (\ref{SR_eqs}) for a quartic potential in the weak dissipation regime, $Q\lesssim 1$. Although there is an ambiguity in the exact choice of the instant $t_i$ (which changes the factor $\sin(2\phi/M)$), we can immediately see that the average eta-parameter correction is suppressed with respect to its oscillation amplitude $\mathcal{A}$ by a factor $(M/M_P)\sqrt{\epsilon_\phi/2}\ll 1$. For the representative parameter choices of our working example in Fig.~1, this yields $\langle \Delta\eta_\phi\rangle\lesssim 10^{-3}$ at horizon-crossing, whereas $\mathcal{A}\simeq 16$. Only at the end of inflation, when $\epsilon_\phi\sim 1+Q$, do we reach $\langle \Delta\eta_\phi\rangle\sim 1+Q$, which explains why the oscillatory thermal corrections to the inflaton's mass do not significantly modify the slow-roll dynamics.

The smallness of the average corrections to the slow-roll parameters when CMB scales exit the horizon also suggests that they should have a negligible effect on the amplitude and shape of the primordial curvature power spectrum on large scales, yielding irrelevant modifications e.g.~to its spectral index or tensor-to-scalar ratio. However, one needs to examine the dynamics of inflaton perturbations more carefully, since as we will discuss in the next section oscillatory corrections to the inflaton's mass may lead to the occurrence of a parametric resonance.


\section{Warm Parametric Resonance}

The dynamics of inflaton perturbations is governed by the Langevin-like equation (\ref{perturbations1}). Since observationally consistent models exhibit weak dissipation when CMB scales become super-horizon, $Q_*\ll 1$, we will ignore dissipative friction in our discussion in this section. We will also begin by considering the solutions of the homogeneous equation, without the noise term $\xi_k$, since these are required to find the full solution of the inhomogeneous equation. We need, however, to take into account the effects of thermal corrections to the inflaton mass, given the large amplitude of their oscillations discussed above, and without performing any averaging procedure. Given that the background field follows a slow-roll trajectory, as we have shown in the previous section, we may neglect its zero-temperature mass, thus yielding:
\begin{align}
\delta\ddot{\phi}_k+3H\delta\dot{\phi}_k+\Big(\frac{k^2}{a^2}-\frac{g^2}{6}T^2\cos(\Omega t+\alpha)\Big)\delta\phi_k=0~.
\label{pertur1}
\end{align}
Let us consider the re-scaled field modes $X_k=a^{3/2}\delta\phi_k$ and change variable to $2z =  \Omega t+\alpha$. We then end up with an equation for a harmonic oscillator with a time-dependent frequency:
\begin{align} \label{X_eq}
\partial_z^2X_k(z)+\tilde{\omega}_k(z)^2X_k(z)=0~,
\end{align}
where
\begin{align}
\tilde{\omega}_k^2(z) = \frac{4k^2}{\Omega^2 a^2}-{2\over3}\frac{g^2T^2}{\Omega^2}\cos(2z)-9\frac{H^2}{\Omega^2}+6\epsilon_\phi \frac{H^2}{\Omega^2}.
\end{align}
Since $H/\Omega \ll 1$ as shown above, we may neglect the last two terms in the previous expression. Defining $A_k(z) = 4k^2/(\Omega^2 a^2)$ and $q = g^2T^2/3\Omega^2$, we then find a Mathieu-like equation:
\begin{align}
\partial_z^2 X_k(z) +\Big(A_k(z)-2q\cos(2z)\Big)X_k(z)=0. \label{mathieu1}
\end{align}
In the Mathieu equation, both $A_k$ and $q$ are time-independent parameters. Despite the fact that in the slow-roll regime we may take $q$ as approximately constant, the parameter $A_k(z)$ will always vary during inflation. Nevertheless, studying the solutions of the Mathieu equation provides a good starting point to determine the dynamics of the inflaton perturbations.

It is well known from Floquet's theorem  \cite{Kofman:1997yn} that the solutions of the Mathieu equation can be written in the general form
\begin{align}
X_k(z) = e^{\mu_kz}\chi(z),\label{mathieu2}
\end{align}
where $\chi(z)$ is some $\pi-$periodic function and $\mu_{k}$ is the Floquet exponent. Depending on the values of $(A_k,q)$, the Floquet exponent may take real ($\mu_k^2>0$) or imaginary ($\mu_k^2<0$) values, leading to unstable or stable solutions, respectively. In the unstable case, we have what is known as a \textit{parametric resonance}. Let us first note that using the slow-roll equations,
\begin{align}
q = \frac{Q}{16C_R}\left(\frac{gM}{T}\right)^2~,
\end{align}
so that in the high-temperature regime $T\gtrsim gM$ where the dissipative coefficient takes the linear form given in Eq.~(\ref{dissip_coeff}), we generically have $q\ll1$ throughout inflation, even when the strong dissipation regime is attained towards its end. This implies that we are interested in the narrow resonance regime, for which the strongest amplification of field modes occurs in the first resonance band, $1-q\lesssim A_k \lesssim 1+q$, where \cite{Kofman:1997yn}:
\begin{align}
\mu_k \simeq \frac{1}{2}\sqrt{q^2-(A_k-1)^2}~.\label{flo1}
\end{align}
The Floquet exponent thus takes its maximum value for $A_k \simeq 1$, i.e.~for the central mode $k_c=a\Omega/2$, which can be interpreted as the production of a pair of inflaton particles with physical momenta corresponding to half of the inflaton mass's oscillation frequency. Moreover, only modes within a narrow momentum range around $k_c$, $\Delta k \simeq q k_c$, are amplified. Since $\Omega\gg H$, amplification of these modes occurs while they are deep inside the Hubble radius.

To estimate the number of particles produced in each mode, we follow a procedure analogous to \cite{Rosa:2007dr}, noting that due to the time-dependence of $A_k\propto a^{-2}$, each mode spends a limited time inside the first resonance band (and similarly for other bands, but where amplification is less significant). The times at which a given mode enters and exits the first resonance band can be computed by setting $\mu_k=0$, yielding:
\begin{align}
\Delta z = \frac{\Omega}{4H}\log\left(\frac{1+q}{1-q}\right)\simeq q{\Omega\over 2H}.
\end{align}
In terms of cosmological time, we have $\Delta t \simeq qH^{-1}\ll H^{-1}$. Thus, each mode spends much less than an Hubble time inside the resonance band and it is interesting to notice that all modes spend the same time inside the band.

We may then reason as follows. While a mode is outside the resonance band, we have $X_k(z)\simeq X_k^{T=0}(z)$, i.e.~with no amplification with respect to the solution in the absence of thermal corrections to the inflaton's mass. Once a mode enters the resonance band, it is amplified by a factor $e^{\mu_k dz}$ in an infinitesimal period $dz$. We thus estimate the total amplification after a mode has exited the resonance band as:
\begin{align}
X_k \simeq X_k^{T=0}\exp\left(\int_{z_i}^{z_f}\mu_k dz\right). \label{sol1}
\end{align}
where $z_i$ and $z_f$ correspond to the instants at which the mode enters and exits the resonance band, respectively (i.e. the solutions for $\mu_k=0$). Since the mode spends less than a Hubble time inside the resonance band, we may linearize the scale factor for $z_i<z<z_f$:
\begin{align}
a(z) = a_c e^{\beta (z-z_c)}\simeq a_c\left(1+\beta(z-z_c)\right) \label{approx1}~,
\end{align}
where $\beta = 2H/\Omega\ll 1$ and $a_c=2k/\Omega$ is the value of the scale factor at which a given mode is in the centre of the resonance band. This then yields $A_k\simeq 1-2\beta(z-z_c)$ and hence
\begin{align}
\int_{z_i}^{z_f}\mu_kdz =\frac{1}{2}\int_{z_i}^{z_f}\sqrt{q^2-4\beta^2(z-z_c)^2} = \frac{\pi}{16}q^2\frac{\Omega}{H}~.
\end{align}
In Appendix A we compare this estimate of the exponential amplification of the inflaton perturbations with a numerical solution of Eq.~(\ref{X_eq}), showing that although not entirely accurate this is sufficient to determine the parametric regimes where inflaton perturbations may be significantly amplified. In terms of the model parameters, the total amplification exponent can be written as:
\begin{align} \label{exponent}
{\pi\over16}\frac{\Omega}{H}q^2 = {\pi\over 2048 C_R^2}\left({gM\over T}\right)^4\left({M_P\over M}\right)\sqrt{2\epsilon_\phi}{Q^2\over 1+Q}~.
\end{align}
In our working example, we have $q\simeq 1.6\times10^{-4}$ and $\Omega/H\simeq 758$ when CMB scales exit the horizon, so that the exponent is extremely suppressed, $(\pi/16) q^2\Omega/H\simeq 4\times 10^{-6}$. Only at the very end of inflation do we find $\mathcal{O}(1)$ values for this exponent, thus showing that inflaton particle production is not efficient. Eq.~(\ref{exponent}) in fact suggests that only in scenarios where much larger values of $Q$ can be attained at horizon-crossing will the parametric resonance amplify inflaton perturbations significantly.

Although we have not considered the effects of the noise term in this discussion, we note that since the solutions of Eq.~(\ref{pertur1}) are not significantly modified by the parametric resonance, the corresponding Green's function will essentially retain its vacuum form. Since the full solution of the inflaton perturbation mode functions can be obtained through the convolution of the vacuum Green's function and the thermal noise $\xi_k$, we conclude that the spectrum of inflaton (and consequently curvature) perturbations will essentially be the same as in the absence of quadratic thermal corrections to the inflaton's mass.


\section{Discussion and conclusions}

In this work we have drawn a very important conclusion for successful model-building in warm inflation scenarios: thermal corrections to the inflaton mass, if oscillatory in nature (with period smaller than the Hubble time), neither spoil the slow-roll dynamics nor significantly change the primordial spectrum of curvature perturbations. 

At the background level, we have shown that only the averaged value of thermal corrections to the slow-roll parameters (particularly eta) has an impact on the dynamics. Although the average corrections are not exactly zero due to the slow change in their oscillation amplitude $\mathcal{A}\propto (T/H)^2$, they are too suppressed in the slow-roll regime to significantly modify the dynamics.

At the perturbation level, although we have identified the occurrence of parametric resonances that may amplify inflaton perturbations, we have concluded that this is not efficient on CMB scales, unless large values of the dissipation coefficient can be attained already when these become super-horizon. This is essentially due to the (very) narrow character of the parametric resonance if $Q$ is not too large. In warm inflation models with a linear dissipation coefficient $\Upsilon\propto T$ (or, in fact, any positive power of $T$), the growth of thermal inflaton fluctuations resulting from their dynamical interplay with fluctuations in the radiation fluid makes scenarios with $Q_*\gtrsim 1$ inconsistent with data, so narrow resonances are expected to occur in all such cases. We cannot exclude, however, a more efficient resonant particle production in scenarios with e.g. $\Upsilon\propto T^{-1}$ that allow for strong dissipation at CMB horizon-crossing, as for instance the scalar version of the WLI model constructed in \cite{Bastero-Gil:2019gao}.

In our case study where the inflaton interacts dominantly with fermion fields, our conclusions have a significant impact on model-building and its possible embedding within concrete extensions of the Standard Model. It is sufficient to consider a single fermion species coupled to the inflaton field and, moreover, there is consequently no need to impose any additional discrete symmetry to cancel out the leading thermal corrections. We note that parametric resonance may also lead to fermion production \cite{Greene:1998nh}, albeit Pauli's exclusion principle makes this much less efficient than boson production. Given our results for inflaton particle production, we also do not expect resonant fermion production to play a significant role.

Yokoyama and Linde \cite{Yokoyama:1998ju} had postulated that warm inflation was impossible to realize because it was, in their view, very hard to keep the fields coupled to the inflaton light and at the same time avoid large thermal corrections to the inflaton's mass. Our work shows that there is a simple way to circumvent both these issues, even simpler than originally proposed in \cite{Bastero-Gil:2016qru}: simply couple the inflaton to fields in such a way that their mass is an oscillatory (and hence bounded) function of the field value. 

Other authors have recently proposed that warm inflation may also be successfully realized with axion-like fields, which are pseudo-NG bosons \cite{Berghaus:2019whh}. These circumvent the issues raised by Yokoyama and Linde since axion-like fields are endowed with shift symmetries broken only by non-perturbative effects. These symmetries could thus protect the inflaton's mass from large thermal corrections, with dissipation arising from its Chern-Simons coupling to massless gauge fields. However, the authors of \cite{Berghaus:2019whh} have only found observationally consistent scenarios with non-renormalizable forms of the inflation potential or hybrid-like scenarios which explicitly (albeit softly) break the crucial shift symmetry. 

The WLI construction is a more robust setup, since the inflaton is not truly a NG boson but rather a gauge invariant degree of freedom, with an arbitrary form of the scalar potential. In the originally proposed setup \cite{Bastero-Gil:2016qru} the potential is somewhat constrained by the additional discrete symmetry, whereas in our simpler WLI scenario this symmetry need not be imposed. Furthermore, the WLI construction is based on a spontaneously broken gauge symmetry, whereas axion models are based on global symmetries, which quantum gravity is not expected to respect. 

In future, we plan to perform full numerical simulations of the evolution of inflaton and radiation perturbations including the noise term, and which may yield precise predictions for the perturbation spectrum. Although we have shown that oscillatory thermal corrections to the inflaton's mass play a negligible role, we hope that these may reveal potential signatures of this effect that, albeit small, may be within the reach of future CMB missions. We also intend to extend our calculations to a version of the WLI model with the inflaton field coupled to a single other scalar field, and determine whether in this case, where strong dissipation is observationally viable, efficient parametric resonances may develop. 

Our analysis has given a much better insight into the true challenges of realizing warm inflation in quantum field theory, and we hope that it may motivate the development of other novel, robust ways, to implement this appealing inflationary paradigm.

\begin{acknowledgments}
We would like to thank Mar Bastero-Gil for useful discussions and for reading our manuscript. J.G.R.~is also indebted to Arjun Berera and Rudnei Ramos for many fruitful discussions on this topic over the past few years.  P.B.F.~is supported by the FCT fellowship SFRH/BD/151475/2021. This work was supported by the CFisUC strategic project No.UID/FIS/04564/2020 and the FCT research grant no. CERN/FIS-PAR/0027/2021.
\end{acknowledgments}

\appendix
\section{Numerical evolution of inflaton perturbations}

In order to validate our analytical estimate of the resonant amplification of inflaton fluctuations, we have solved Eq.~(\ref{X_eq}) numerically starting from vacuum initial conditions, computing in particular the occupation number of each comoving momentum mode $k$, given by \cite{Kofman:1997yn}:
\begin{align}
n_k = \frac{\omega_k}{2}\left(|X_k|^2+\frac{|\dot{X}_k|^2}{\omega_k^2}\right)-\frac{1}{2}~.
\end{align}
For convenience we express our results in terms of the variable $Z_k = n_k+1/2$. The expectation is that $Z_k$ remains constant, or at most exhibits an oscillatory behaviour, while the mode is outside the resonance band, being amplified as $e^{\int_{z_i}^z\mu_k dz}$ while the mode is within the resonance band. This is indeed what we obtain, as shown in Fig.~3 both for the realistic parameter choices of our working example in Figs.~1 and 2, and an alternative (unrealistic) choice of parameters yielding a significant amplification of the modes.

\begin{figure}[!h] \label{fig3}
\centering
\includegraphics[scale = 0.38]{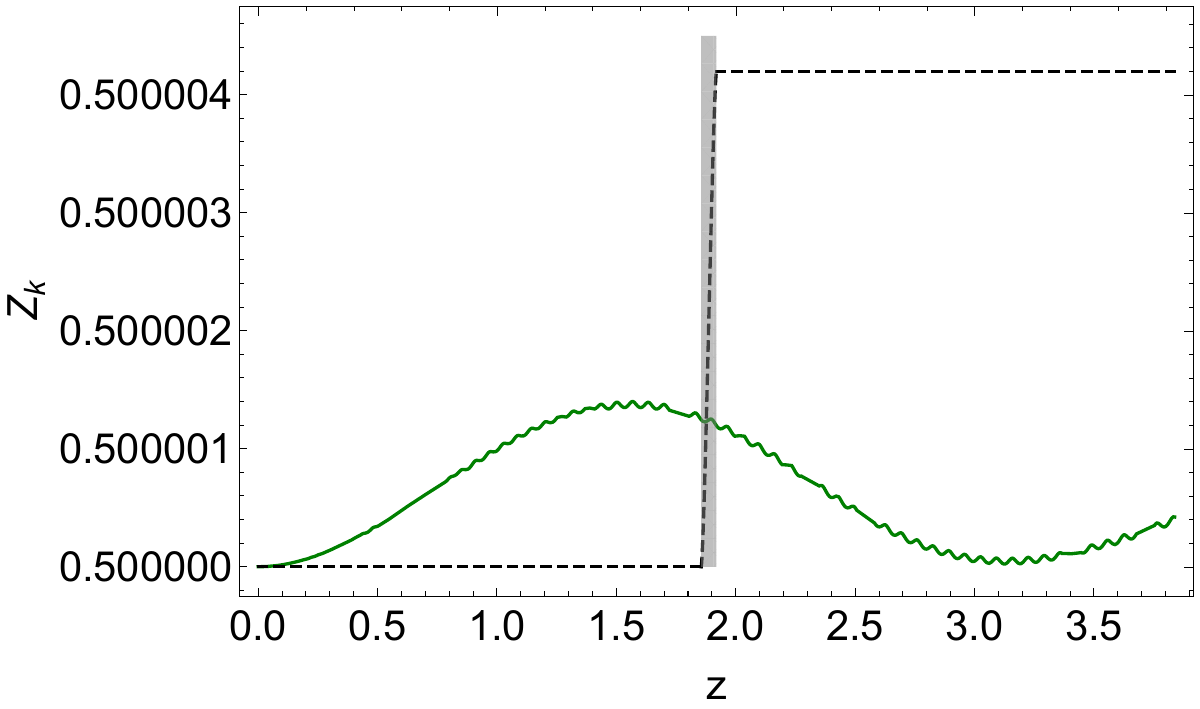}
\includegraphics[scale = 0.35]{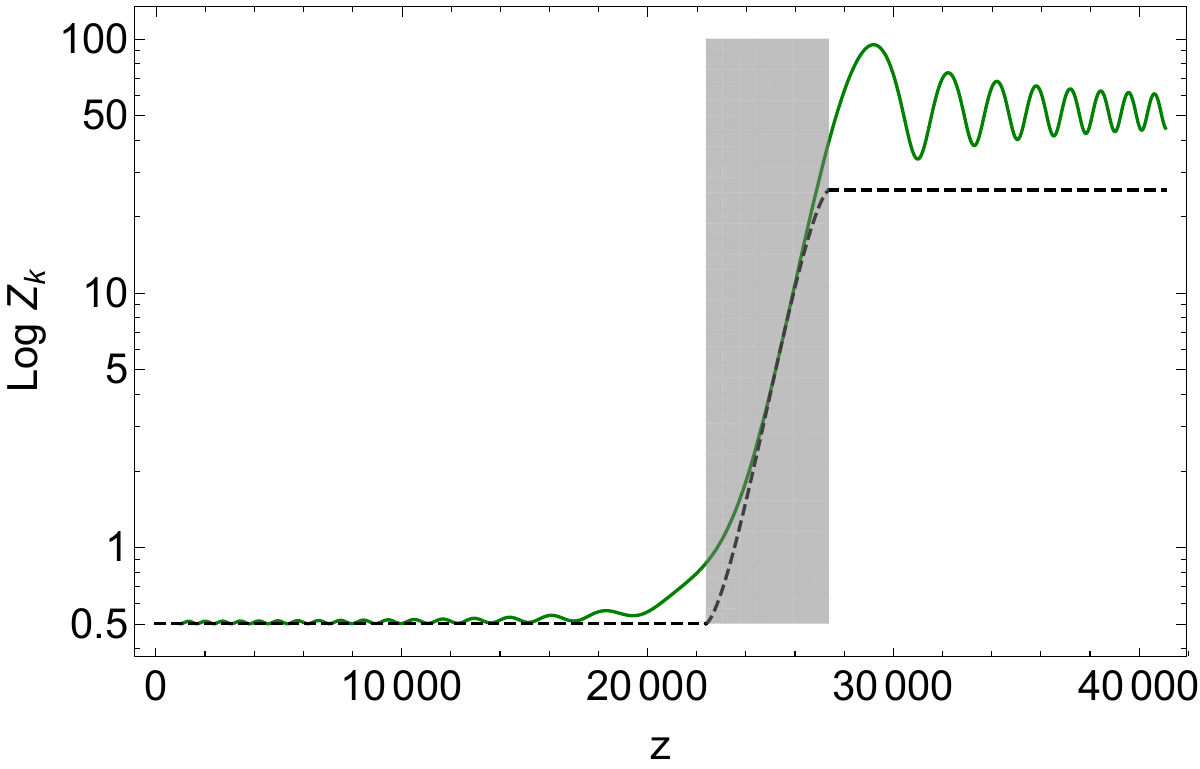}
\caption{Numerical evolution (green curve) of the particle number for an illustrative mode $k$ with the parameter choices and horizon-crossing conditions given in Fig.~1, corresponding to $q=1.6\times10^{-4}$ and $\Omega/H\simeq 758$ (left); and with an unrealistic choice of parameters $q=10^{-3}$ and $\Omega/H =10^7$ (right). The black dashed curve gives the approximate analytical solution in each case. The shaded region corresponds to the period the mode is inside the first resonance band.}
\end{figure}

Our analytical estimate is not fully accurate, particularly in the (unrealistic) case with a large mode amplification, where we underestimate the final particle number, essentially due to the mode's behaviour when exiting the resonance band, which is not fully captured by the integrated effect of the Floquet exponent (while the behaviour of the mode while inside the resonance band is accurately described). Of course for large values of the exponent any innaccuracy in its determination may have an exponentially large effect, but since this is not a realistic case an improvement of our estimate is beyond the scope of this work.

On the contrary, we slightly overestimate the particle number in the realistic scenario, although its order of magnitude is correctly captured by our analytical approximation. We nevertheless clearly see that when the resonance is too narrow and the mode spends only a very short period inside the resonance band, essentially no particle production occurs and the mode remains in the vacuum state.

\end{document}